\documentclass[11pt]{article}
\usepackage[T1]{fontenc}
\usepackage[utf8]{inputenc}
\usepackage[margin=1in]{geometry}
\usepackage{graphicx,amsmath,booktabs,tabularx,placeins,microtype}
\usepackage[hidelinks]{hyperref}
\hypersetup{pdftitle={Measuring Brand and Source Discovery under Repeated LLM Queries: A Finite-Sample Audit},pdfauthor={Dmitrij Zatuchin},pdfsubject={Research manuscript, 11 September 2026}}
\usepackage{parskip}
\usepackage{fancyhdr}
\title{\bfseries Measuring Brand and Source Discovery under Repeated LLM Queries:\\A Finite-Sample Audit}
\author{Dmitrij \.Zatuchin\\[3pt]
\small Estonian Entrepreneurship University of Applied Sciences, Tallinn, Estonia\\
\small Rankfor.AI, Tallinn, Estonia\\
\small \texttt{dmitrij.zatuchin@eek.ee}}
\date{11 September 2026}
\begin{document}
\maketitle
\begin{abstract}
Repeated-query audits must distinguish recovery of a collected set from completeness of possible outputs. We apply sample-based rarefaction to 4,500 responses from 50 buying questions, six configurations and 15 calls per cell. Historical-dictionary median ten-call recovery of the observed 15-call set ranges from 92.6\% to 95.2\%; re-adjudicating all 45,683 candidate strings changes this range to 89.5\%--94.7\%. Two blinded Gemini 3.1 Pro annotation roles assessed 600 complete answers, yielding micro F1 of 0.908 for canonical-name agreement and 0.975 for span-overlap agreement. This is AI-based evidence, without a human reference study. A separate matched roster analysis of 3,750 records per wave gives median single-call recovery of the observed five-call set of 80.0\%--92.5\% in February and 90.0\%--100.0\% in September, with question-subset dependence. Source accumulation also changes when API-returned hosts are restricted to those referenced by answer citation markers. These findings show that recovery percentages depend on extraction, question selection and the finite reference collection. They support explicit measurement definitions and sensitivity analyses, without establishing exhaustive repertoires, causal retrieval effects or a universal stopping rule.
\end{abstract}

\section{Introduction}
An audit of a language model often begins with a set of questions and a budget of repeated calls. The resulting report may list brands mentioned in the answers or websites returned as sources. Each additional call can add an organization or source that previous calls missed. Consequently, the observed set depends on the number of calls, the deployment configuration and the extraction procedure.

Repeated-output variability is documented even under settings intended to be deterministic \cite{atil}. Research on generative search also distinguishes accurate, comprehensive citation of an answer from the mere presence of citation links \cite{liu,gao}. Source-selection research describes differences between providers \cite{yang}. These questions motivate a separate measurement problem: how much of the set found across a specified collection would a smaller repeated-query sample have recovered?

This paper applies established incidence-based estimators to that problem \cite{chao,colwell}. Its contribution is an empirical audit of the measurement process: it separates observed-set recovery from estimated unseen richness, contrasts organization mentions with source-host records, and tests how data processing changes apparent completeness. The estimators themselves are established methods. We make no claim to a new richness estimator or an exhaustive inventory of all possible model outputs.

We ask three questions. First, how does the observed organization set accumulate across repeated calls? Second, how does source accumulation depend on the definition of a source observation? Third, which apparent completeness conclusions survive an audit of missing responses, vocabulary restrictions and estimator interpretation? The September 4 six-engine experiment is the main analysis. A matched February/September comparison evaluates portability under a fixed closed-roster instrument. An audit of historical processing and four earlier deep cells provide additional checks. The panels measure different item universes and are not pooled into a single recovery estimate.

\section{Data and measurement}
\subsection{September experiment}
The collection contains 50 English buying questions, selected deterministically as the first ten questions in each of five existing industry banks: software as a service, consulting, financial technology, e-commerce and health technology. The selected questions contain no aliases from the study's historical brand roster. Examples include ``What is the best CRM for small businesses?'' and ``What is the best telemedicine app?'' This is a designed question panel, not a probability sample of user demand.

Each question was sent 15 times in a fresh conversation to six API configurations, yielding 300 cells and 4,500 responses. All scheduled cells have 15 distinct iteration identifiers, non-empty answer text and no recorded error. Completion timestamps run from 3 September 2026 at 22:56:25 UTC to 4 September at 05:40:50 UTC; the collection falls on 4 September in Tallinn. Task order was shuffled with seed 2026 and executed with eight workers. Iteration identifiers therefore do not encode chronological completion order.

The stored engine identifiers are \texttt{gpt-5.6-luna}, \texttt{claude-sonnet-5}, \texttt{gemini-3.7-flash}, \texttt{grok-4.5}, \texttt{mistral-large} and \texttt{sonar}. The returned model-version strings repeat those names, except that Mistral is recorded as \texttt{mistral-large-latest}. These are the recorded identifiers; they do not establish immutable backend snapshots. The collector used provider-default sampling parameters. The Anthropic request explicitly capped output at 1,024 tokens; other providers' requests did not impose the same cap. Sonar supplied search-related source metadata; the other five configurations were called without an explicit search tool. Retrieval was not randomized within models, so differences between engines cannot isolate a retrieval effect.

\subsection{Organization extraction}
The stored extraction pools capitalized candidate strings by question across engines. A Gemini classifier, called at temperature zero, maps retained strings to organization names under a rule requiring an organization to offer the queried category. Product names may be mapped to an organization, and documented subsidiaries may be consolidated with a parent. The dictionary contains 1,470 distinct canonical strings across all questions and 2,280 question-specific canonical entries. These are classifier-assigned names, not an independently verified census of organizations.

The implementation applies a frequency filter before classification: a candidate is eligible if it occurs at least twice in the pooled candidate record, contains at least two whitespace-separated words, or contains an ampersand or digit. The audit found 10,351 question-specific candidate strings that meet this exclusion predicate. Two, Wisp and ICF in their respective questions, nevertheless occur in the saved dictionary; the old output parser accepted returned candidates from the entire pool rather than only the submitted chunk. The 10,351 count is therefore not a count of guaranteed dictionary omissions. Many are ordinary words, but the filter can remove relevant rare organizations. For example, the small-business CRM response from GPT, iteration 3, names GoHighLevel in a comparison table; that single-word candidate is filtered out and has no entry in the question's retained dictionary. The extraction therefore cannot be treated as unrestricted open recognition.

Retained aliases are matched case-insensitively in the full response and deduplicated within a run. All eligible mentions count; the procedure does not distinguish a positive recommendation from a comparison, integration reference or negative mention. We consequently refer to organization mentions or extracted brands. A sensitivity check replacing ASCII alphanumeric boundaries with Unicode word boundaries changed no run-level sets in this English panel. This check does not validate semantic eligibility, alias mapping or extraction recall. The results below remain conditional on the specified vocabulary. A separate blinded AI audit assesses dictionary agreement with complete-answer annotations; no human reference study was conducted.

\subsection{Candidate re-adjudication}
On 11 September, we passed all 45,683 stored question-specific candidate strings to the original Gemini 3.7 Flash classifier without the frequency eligibility filter. The original organization eligibility rule was retained. Each candidate was accompanied by one literal answer excerpt and the buying question; engine, date and frequency metadata were withheld. Answer text could still reveal its source. Lexically sorted chunks contained at most 200 candidates, with temperature zero and an output cap of 8,192 tokens. Every input index required an explicit keep, reject or uncertain decision. Missing indices, invalid JSON and failed requests could not become empty accepted lists. All 255 chunks completed without a failed attempt.

This is a sensitivity analysis after inspection of historical results. We retain three dictionaries: the historical baseline; an add-only arm preserving baseline entries and adding absent, formerly filtered strings now classified as eligible; and full re-adjudication. Case-folded canonical names and exact old-alias matches align certain naming variants, with each change logged. These operations are not comprehensive entity resolution. The same case-insensitive alias matcher scores the same answers in all three arms. Full re-adjudication changes context and chunk composition as well as candidate eligibility, so its whole effect cannot be assigned to filter removal.

The classifier kept 6,867 strings, rejected 38,815 and marked one uncertain. The uncertain candidate, Zavala in the telehealth-platform question, remains unresolved and is excluded from the accepted dictionary. The capitalization-based candidate generator and automated eligibility decisions can still produce false negatives or false positives. We sampled 600 complete answers, two per cell with seed 20260911, for the additional AI audit described below. Prepared human-review packets remain unannotated.

\subsection{Blinded AI extraction audit}
We assessed the candidate dictionaries against complete-answer annotations from Gemini 3.1 Pro Preview \cite{gemini31}. Two role prompts applied the same eligibility rubric in fresh stateless API requests: an organization-mention annotator (A) and a replication auditor (B). Each reviewed all 600 sampled answers, in separately shuffled batches of five (seeds 311 and 312). Inputs contained the answer ID, buying question and full answer. The production dictionaries, manuscript claims and explicit engine/date metadata were withheld; answer text could still reveal its source. These were AI review roles, not human participants.

A third role (C) reviewed answers where normalized canonical sets differed or either reviewer recorded uncertainty. C received the full answer and randomly ordered A/B annotations, without production dictionary labels. Equal A/B sets with no recorded uncertainty were retained, combining their literal surfaces. The resulting reference is model-generated; agreeing passes may share errors. All roles used \texttt{gemini-3.1-pro-preview}, temperature 1, medium thinking and a 16,384-token output cap. Search grounding was disabled, and parent mappings relying on model knowledge were recorded as such rather than described as externally verified. Returned model identifiers were preserved.

The local protocol and sample were frozen before annotation. Six additional answers outside the sample piloted the schema and rubric application. Every final record required complete answer-ID coverage, a completed generation and exact literal surfaces. Invalid outputs were retained as failures and retried. Batches that exhausted three attempts received a documented extension of up to three format-repair attempts, with the same role, rubric and answers plus explicit JSON and contiguous-span instructions. No failed batch was dropped or scored as empty.

We report A/B canonical-set agreement and micro F1 after Unicode normalization, case folding and punctuation removal. A separate maximum one-to-one span-overlap match allows different canonical labels to agree on the same literal name occurrence. Against the reconciled AI reference, each dictionary receives strict-name and span-overlap precision, recall and F1. The three scored dictionaries were retained unchanged. These measures describe dictionary-to-model-reference agreement, not accuracy against human ground truth. Model-recorded uncertainty flags, including proposed exclusions and mapping cautions, are retained separately; only accepted organization lists enter the scores. Literal strings are matched at all occurrences because reviewers did not supply disambiguated offsets. Shared short names and nested names can inflate span agreement, which is a relaxed matching sensitivity rather than proof of entity identity. Because only two answers per 15-answer cell are annotated, this audit cannot recompute complete-panel accumulation curves. It does not validate the separate matched-wave roster matcher.

\subsection{Matched five-call collections}
A separate comparison uses 3,750 February 23--24 records and 3,750 September 9--10 records. All 250 question texts, three nominal requested configurations and five iteration identifiers match. Requested configurations are GPT-5.2, Gemini 3 Flash preview and Sonar Pro. September records return \texttt{gpt-5.2-2025-12-11}, \texttt{gemini-3-flash-preview} and \texttt{sonar-pro}. February lacks returned version strings. Nominal collection settings use temperature 0.3 and output limits of 1,024 tokens, with provider-specific token semantics. Collection moved from Google Colab, with region unrecorded, to the author's northern-European workstation; this prevents attribution solely to elapsed time.

The same existing guarded matcher, 77 aliases and 50 predefined brands are applied to both waves. The matcher uses word boundaries and additional capitalization or context guards for ambiguous aliases. The primary instrument counts only the ten roster brands assigned to the question's industry; a sensitivity counts all 50, including off-industry mentions. Thus this comparison measures a closed roster, distinct from the six-engine panel's broader candidate extraction. The recomputed February presence pairs match all 10,463 previously corrected response/brand pairs exactly. The guarded matcher was not independently human validated. The 600-answer AI audit concerns the broader candidate extraction and supplies no validation of this separate roster instrument.

There are 136 blank February answers and no blank September answers. Primary analysis retains the 707 question/system pairs with five non-empty answers in both waves. Non-empty answers with zero matched brands remain sampling units. Recovery ratios are undefined for empty unions and are summarized over pairs with non-empty unions in both waves. A separate sensitivity retains all 750 scheduled cell pairs, assigning an empty observed set to each blank answer for a call-budget estimand. Neither convention establishes why February answers were blank.

The local plan and input hashes were frozen at 03:14:36 UTC on 11 September before calculating this task's September 9--10 brand outcomes. Collection had already occurred and historical outcomes were known: this is a local analysis specification, not public preregistration. It fixes the primary comparison and sensitivities, including comparison-prompt exclusion (indices 15--24) and the 50 prompts shared with September 4 (indices 0--9). The five-call records lack the source arrays required for a source-host comparison and cannot test ten-call recovery.

\subsection{Source observations}
For Sonar, the primary source measure includes every URL in the stored API citation list. We parse the URL host, lowercase it, remove a leading \texttt{www.}, and deduplicate hosts within a run. Subdomains remain distinct; this is a host-level measure, not registrable-domain or publisher ownership consolidation. Returned source metadata can include links that are not referenced by the answer text.

We therefore add a separate sensitivity analysis. Numeric markers of the form \texttt{[n]} in the answer are mapped to the corresponding one-based entries in the stored citation list. The sensitivity set contains the hosts of those entries. This measures visible marker use under that mapping convention. It does not assess whether the source supports the surrounding claim, whether an answer omits a necessary citation, or whether the system used an unreported source. These are distinct dimensions of citation quality \cite{liu,gao}.

\section{Analysis}
Let $B_i$ be the set of extracted items in run $i$, $n$ the number of retained runs, $S_{\mathrm{obs}}=|\bigcup_{i=1}^n B_i|$, and $f_j$ the number of runs containing item $j$. Exact sample-based rarefaction gives the expected number of observed items recovered by a uniformly selected subset of $k$ runs:
\begin{equation}
 A(k)=\sum_{j=1}^{S_{\mathrm{obs}}}\left[1-\frac{\binom{n-f_j}{k}}{\binom{n}{k}}\right],\qquad 1\leq k\leq n,
\end{equation}
where the numerator is zero when $k>n-f_j$. This is a finite-sample interpolation calculation. It does not require identifying an unobserved population to describe the collected runs. Interpreting it as a guide to future calls additionally requires a sufficiently stable deployment and sampling process.

We report $A(1)/A(5)$ and $A(10)/A(15)$ in the September 4 experiment. The first ratio compares expected one-run and five-run richness within the same 15-run reference collection. The second gives expected recovery of the observed 15-run set. Neither ratio is the fraction of every organization the engine could ever mention. We summarize cells by engine using medians; no hypothesis test or population-level confidence claim is attached to this deterministic panel.

In the matched five-call comparison, $A(5)=S_{\mathrm{obs}}$. We report each wave's median $A(1)/A(5)$ and the median within-pair difference separately; a difference of medians is not the median paired difference. The local descriptive criterion for ``about 80\%'' is a September median in $[0.75,0.85]$ per system on the primary paired sample. This convention is not an equivalence test, statistical significance threshold or evidence that the older, differently extracted pooled result should match it.

Let $Q_1$ and $Q_2$ count items found in exactly one and two runs. The terminal increment of the rarefaction curve is
\begin{equation}
 A(n)-A(n-1)=Q_1/n.
\end{equation}
Thus $Q_1>0$ identifies a positive expected increment when expanding a randomly selected $(n-1)$-run subset to the full observed sample. It does not identify the yield of the final chronological call. We calculate that yield separately by sorting on completion timestamps.

For comparison with the previous analysis, we also reproduce its Chao2 convention:
\begin{equation}
 \widehat S_{\mathrm{Chao2}}=S_{\mathrm{obs}}+\frac{n-1}{n}
 \begin{cases}
 Q_1^2/(2Q_2),&Q_2>0,\\
 Q_1(Q_1-1)/2,&Q_2=0.
 \end{cases}
\end{equation}
Chao2 is a lower-bound richness estimator under its sampling assumptions \cite{chao,colwell,chaochiu}. The ratio $S_{\mathrm{obs}}/\widehat S_{\mathrm{Chao2}}$ can be optimistic about completeness because the denominator is not a known total or an upper bound. In particular, this convention returns $\widehat S=S_{\mathrm{obs}}$ when $Q_1=1$ and $Q_2=0$, despite the observed singleton. We treat it as a diagnostic and do not use it to certify a stopping rule. Independent subset enumeration on a small incidence example verifies the rarefaction implementation; all 300 previously stored September $S_{\mathrm{obs}}$ and $Q_1$ values reproduce, as do the Chao2 estimates to their saved precision.

\section{Results}
\subsection{Historical-dictionary results in the 15-call panel}
\begin{figure}[htbp]
\centering\includegraphics[width=\textwidth]{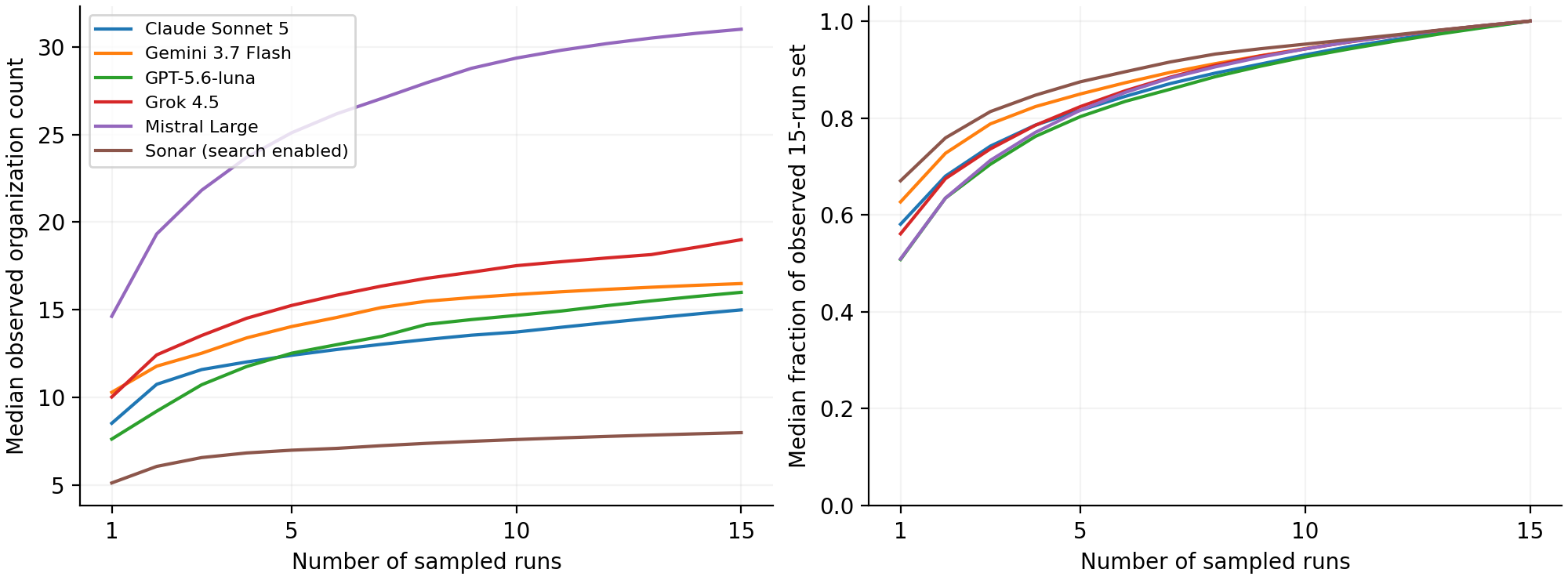}
\caption{Expected organization discovery, conditional on the stored candidate-based extraction. Left: median distinct counts. Right: median recovery relative to each cell's observed 15-run set. Curves describe the collected panel; they do not identify the fraction of the entire possible output repertoire.}\label{fig:brands}
\end{figure}

Table~\ref{tab:engines} summarizes the historical-dictionary September 4 results. The observed organization set has a median size of 8 for Sonar and 15 to 31 for the other configurations. A single sampled run recovers 62.2\% to 76.8\% of the expected five-run set. After ten sampled runs, median recovery of the observed 15-run set is 92.6\% to 95.2\% across all six configurations (Figure~\ref{fig:brands}).

\begin{table}[htbp]
\centering\small
\caption{Historical dictionary, September 4 experiment: 50 cells per engine and 15 runs per cell. All ratios and set sizes are medians over cells. $Q_1>0$ counts cells with at least one singleton.}\label{tab:engines}
\begin{tabular}{@{}lrrrr@{}}\toprule
Engine & $S_{\mathrm{obs}}$ & $Q_1>0$ & $A(1)/A(5)$ & $A(10)/A(15)$\\\midrule
Claude Sonnet 5 & 15 & 45/50 & 0.708 & 0.930\\
Gemini 3.7 Flash & 16.5 & 45/50 & 0.724 & 0.943\\
GPT-5.6-luna & 16 & 46/50 & 0.647 & 0.926\\
Grok 4.5 & 19 & 43/50 & 0.672 & 0.943\\
Mistral Large & 31 & 46/50 & 0.622 & 0.943\\
Sonar, search enabled & 8 & 32/50 & 0.768 & 0.952\\\bottomrule
\end{tabular}
\end{table}

Singletons remain in 43 to 46 of 50 cells for the five configurations without explicit search, and in 32 of 50 Sonar cells. The corresponding terminal rarefaction increments are positive. In contrast, the last completed call adds an organization absent from previous completed calls in 9 to 11 of 50 cells for each of the five configurations and in 2 of 50 Sonar cells. The difference demonstrates why singleton prevalence must not be described as the percentage of final chronological calls that discover a new brand.

Sonar's median ratio to Chao2 is 1.00, even though 64\% of its cells contain singletons. Twelve Sonar cells have $Q_1=1$, $Q_2=0$ and consequently a zero Chao2 unseen correction. The ratio cannot establish that Sonar's repertoire is exhausted. Across engines, the median question has 38.5 observed organizations in the six-engine union, 15 of which occur in exactly one engine. The best single engine covers a median 83.1\% of that observed union. This union is a property of the panel and extraction procedure, not a market inventory.

Output length also differs. The median of cell-level median character counts is 1,341.5 for Sonar and 8,315.5 for Mistral. These differences, together with unequal output caps and different models, limit explanations based on retrieval. An engine that produces longer answers has more opportunities to mention additional organizations.

\subsection{Sensitivity to candidate eligibility and re-adjudication}
The add-only arm restores 999 question-specific candidate strings, including GoHighLevel in the small-business CRM question. It changes organization sets in 475 of 4,500 answers (10.6\%). Full re-adjudication changes 2,249 answer sets (50.0\%) through additions, removals and canonical mappings. These counts describe changed measurements, not verified improvements in accuracy. A single occurrence in the candidate pool is not necessarily a response-level singleton after alias matching and consolidation.

\begin{table}[htbp]
\centering\small
\caption{Sensitivity on identical September 4 answers. H: historical dictionary; A: add-only restoration; F: full re-adjudication. All entries are medians over 50 cells per configuration.}\label{tab:repair}
\begin{tabular}{@{}lrrr@{}}\toprule
 & \multicolumn{3}{c}{$A(10)/A(15)$}\\
Configuration & H & A & F\\\midrule
GPT-5.6-luna & 0.926 & 0.913 & 0.910\\
Claude Sonnet 5 & 0.930 & 0.921 & 0.920\\
Gemini 3.7 Flash & 0.943 & 0.934 & 0.936\\
Grok 4.5 & 0.943 & 0.926 & 0.923\\
Mistral Large & 0.943 & 0.897 & 0.895\\
Sonar & 0.952 & 0.947 & 0.947\\\bottomrule
\end{tabular}
\end{table}

Median ten/fifteen-call recovery ranges from 89.5\% to 94.7\% under full re-adjudication (Table~\ref{tab:repair}), compared with 92.6\%--95.2\% historically. Mistral shows the largest change: median observed richness is 31 historically, 38 after add-only restoration and 40.5 after full re-adjudication. Its corresponding recovery changes from 94.3\% to 89.7\% and 89.5\%. Full re-adjudication leaves singletons in 33--48 of 50 cells per configuration. High recovery and residual discovery therefore coexist, while the precise recovery percentage remains extraction-sensitive.

\begin{figure}[htbp]
\centering\includegraphics[width=\textwidth]{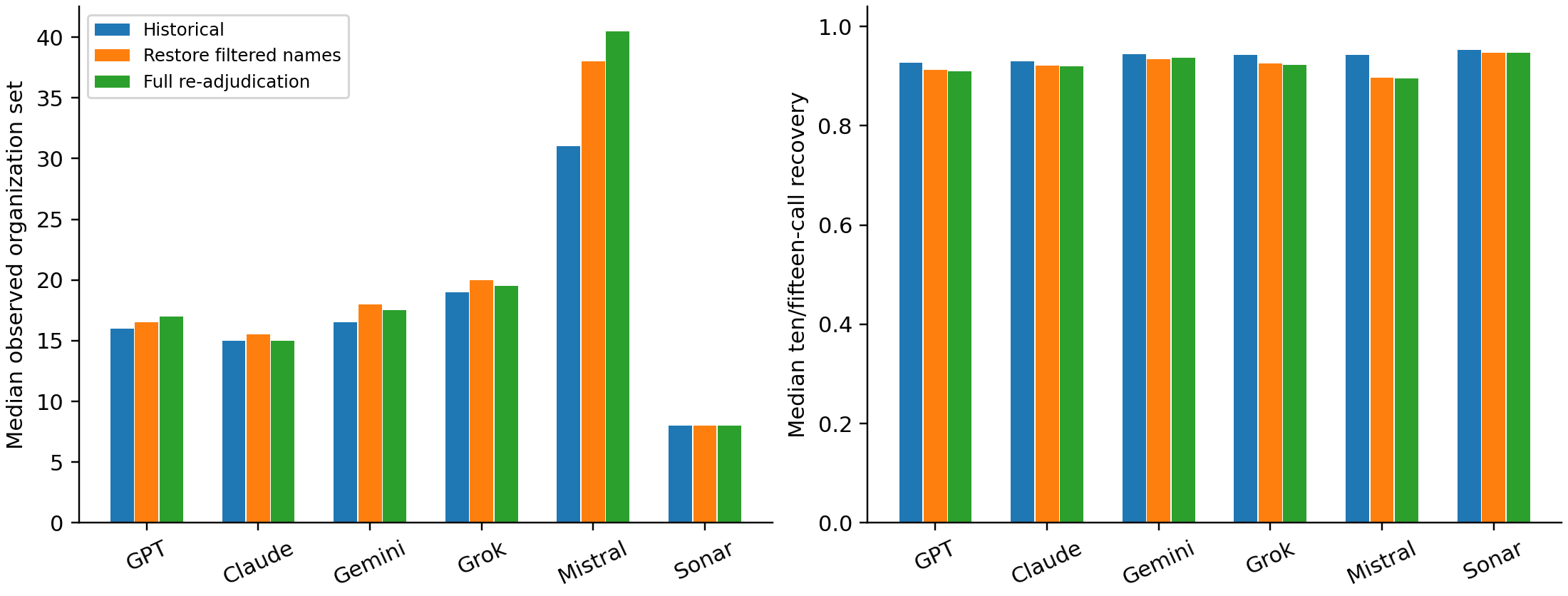}
\caption{Dictionary sensitivity on the same responses. The add-only arm preserves baseline errors; full re-adjudication also changes classifier context and organization mapping. Neither revised arm is an independently validated gold standard.}\label{fig:repair}
\end{figure}
\FloatBarrier

\subsection{Agreement with an AI reference}
Both reviewer roles completed all 600 answers. Their normalized canonical sets were identical on 281/600 answers (46.8\%); micro F1 was 0.908 for strict canonical names and 0.975 for one-to-one span overlap. C adjudicated 400 answers with differences or uncertainty. The AI reference contains 6,019 answer/organization presences, with 579 model-flagged entries across 214 answers retained separately. These flags include proposed exclusions and mapping questions, rather than representing a count of unique unresolved organizations. The distinction between canonical and span agreement matters: different parent or corporate names can refer to overlapping surface text without proving that either mapping is correct.

\begin{table}[htbp]
\centering\small
\caption{Agreement with the AI reference on the same 600 answers. P: precision; R: recall; F1: harmonic mean. These are conditional model-reference scores, not human-ground-truth accuracy.}\label{tab:ai}
\begin{tabular}{@{}lrrrrrr@{}}\toprule
& \multicolumn{3}{c}{Strict canonical names} & \multicolumn{3}{c}{Literal span overlap}\\
Dictionary & P & R & F1 & P & R & F1\\\midrule
Historical & 0.848 & 0.865 & 0.857 & 0.918 & 0.936 & 0.927\\
Add-only & 0.847 & 0.877 & 0.861 & 0.915 & 0.947 & 0.931\\
Full re-adjudication & 0.844 & 0.883 & 0.863 & 0.916 & 0.959 & 0.937\\
\bottomrule
\end{tabular}
\end{table}

Table~\ref{tab:ai} quantifies agreement under each matching convention. Full re-adjudication increases strict-name recall against this reference from 0.865 to 0.883 and span-overlap recall from 0.936 to 0.959, with small decreases in precision. This is a comparison against the same AI reference, not an independent accuracy estimate. Dictionary-only and reference-only entities may reflect extraction omissions, eligibility decisions or organization mappings. The reference is not an independently verified census, and a higher agreement score alone cannot establish that an extraction arm is more accurate. Per-answer disagreements, unresolved cases and per-engine summaries are supplied with the audit records. The observed rarefaction results remain conditional on their respective dictionaries.
\FloatBarrier

\subsection{Matched February/September comparison}
Table~\ref{tab:waves} reports the primary own-industry roster comparison. September medians are 90\%, 90\% and 100\%, so none lies in the locally specified 75--85\% descriptive band. This does not establish a pure time effect or statistically reject the earlier pooled result, which used a different extraction instrument. It does show that ``about 80\%'' is not a portable per-system description under the present instrument.

\begin{table}[htbp]
\centering\small
\caption{Paired non-empty five-call cells, own-industry roster. Defined pairs have non-empty unions in both waves. Recovery medians use these same pairs. Changes are median within-pair differences, in percentage points.}\label{tab:waves}
\begin{tabular}{@{}lrrrrr@{}}\toprule
Configuration & Complete & Defined & February & September & Change\\\midrule
GPT-5.2 & 207 & 203 & 0.900 & 0.900 & 0.0\\
Gemini 3 Flash preview & 250 & 237 & 0.925 & 0.900 & 0.0\\
Sonar Pro & 250 & 221 & 0.800 & 1.000 & +5.7\\\bottomrule
\end{tabular}
\end{table}

Sonar's difference of medians is 20 percentage points, but its median paired change is 5.7 points. These are distinct summaries. Median observed roster richness is three for GPT and two for Gemini in both waves, and three then two for Sonar. A 100\% median describes recovery of these small collected roster sets; it does not imply comprehensive organization discovery or identical answers.

\begin{figure}[htbp]
\centering\includegraphics[width=\textwidth]{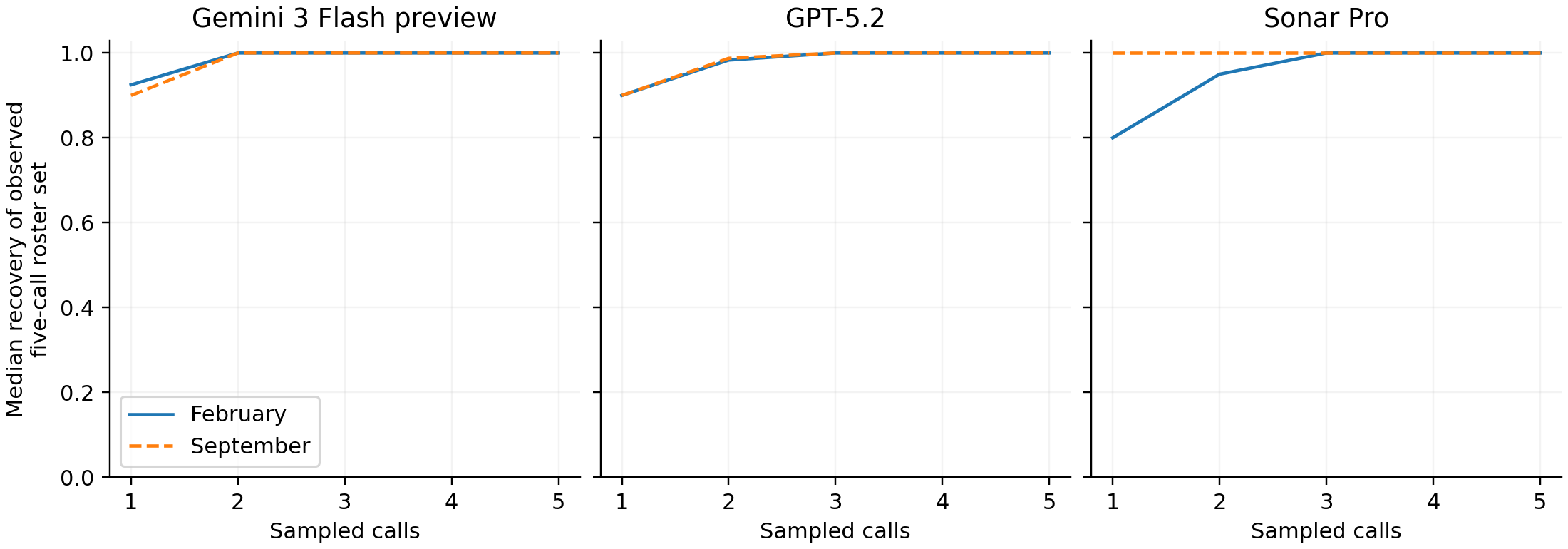}
\caption{Median finite-sample recovery of the observed five-call own-industry roster set, over the same defined pairs in each wave. A flat median does not mean every question has a flat curve.}\label{fig:waves}
\end{figure}

Results depend on the question subset. On the 50 shared prompts, February/September medians are 80.0\%/88.9\% for GPT (49 defined pairs), 80.0\%/80.0\% for Gemini (49) and 80.0\%/100.0\% for Sonar (45). Excluding comparison prompts gives 86.7\%/88.0\%, 82.9\%/85.0\% and 77.5\%/97.8\%, respectively. This subset is not certified brand-free. Including all 50 roster brands, irrespective of industry, gives complete-pair medians of 83.3\%/84.4\%, 86.7\%/86.7\% and 76.3\%/100.0\%. The accompanying tables retain all denominator and blank-answer sensitivities.
\FloatBarrier

\subsection{Source discovery changes with source definition}
Using all API-returned source hosts, Sonar has a median of 21 hosts across 15 runs, with singletons in 22 of 50 cells. The median $A(1)/A(15)$ is 88.0\%. Using hosts attached to numeric citation markers, the median observed set is 13 hosts, singletons occur in 39 of 50 cells and median $A(1)/A(15)$ falls to 59.5\%. Repeated API source lists and variation in which entries an answer references produce different discovery curves. A reported citation surface must identify which record it counts.

\begin{figure}[htbp]
\centering\includegraphics[width=\textwidth]{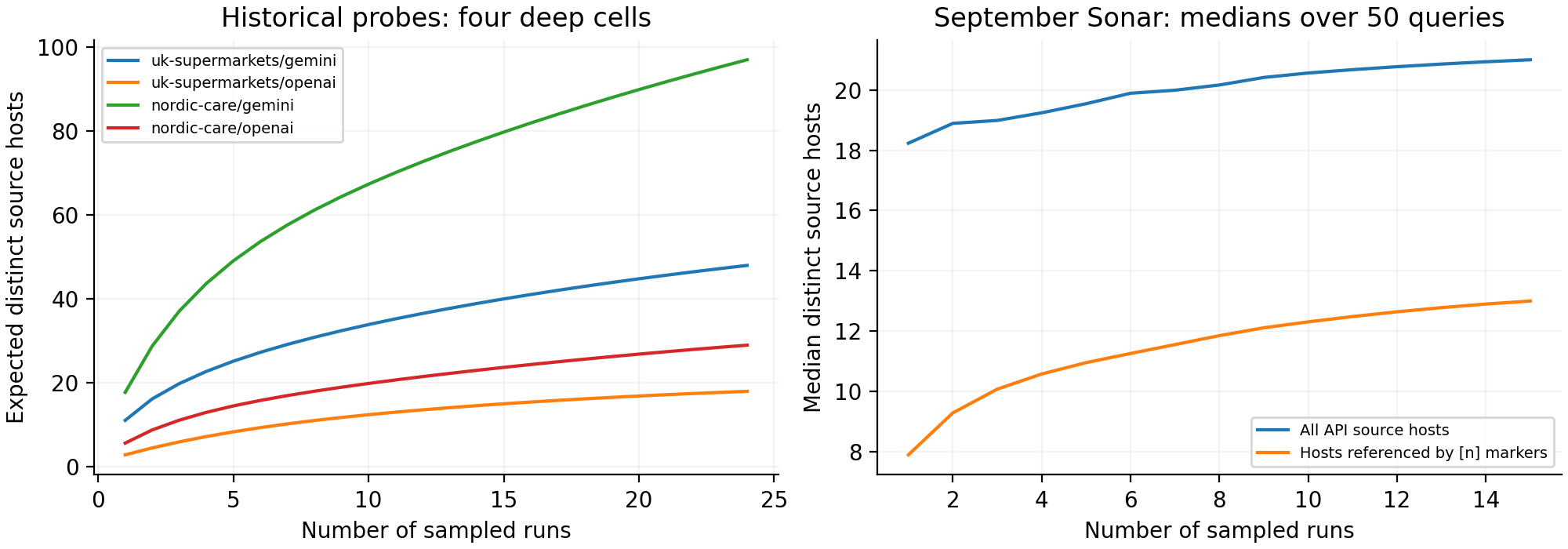}
\caption{Source-host accumulation. Left: four earlier 24-run cells using their recorded source sets. Right: the September Sonar panel, distinguishing all returned source metadata from hosts referenced by numeric markers in the answer. Each line represents the indicated operational definition.}\label{fig:domains}
\end{figure}
\FloatBarrier

\section{Interpretation and limitations}
Across the September 4 configurations, repeated queries expand the extracted organization set, ten runs recover most of the observed 15-run set, and rare items can remain even when normalized recovery is high. The precise numerical range changes after candidate re-adjudication, while the separate matched-wave analysis also shows dependence on question selection and roster definition. Both facts can hold because an accumulation curve may have a small positive terminal increment. A visually flat curve alone is inadequate evidence for exhaustive discovery.

The practical contribution is a reporting procedure with an explicit reference set. An auditor can state the query panel and call count, define the extraction universe, plot $A(k)$, and report the terminal increment alongside observed-set recovery. A report saying that ten runs recovered approximately 93\% of its 15-run reference set is interpretable. A claim that ten runs recovered 93\% of everything an engine could recommend requires additional evidence. Sampling completeness in the sense of probability mass also differs from the proportion of distinct item types recovered \cite{chaochiu}.

Brand and source results answer different questions. Organization mentions describe the extracted answer content. API source hosts describe returned metadata, and marker-linked hosts describe a subset visibly referenced in the answer. None directly measures verified evidential support. An audit concerned with the reliability of recommendations must add claim-level assessment \cite{liu,gao}. An audit concerned with recommendation frequency must also estimate incidence probabilities with an appropriate repeated-measurement design \cite{dice}; discovering a set does not settle how often its members appear.

Several limitations restrict the empirical interpretation. First, vocabulary construction is selective and has a documented rare-name omission. Capitalization, pooled-frequency filtering, automated adjudication and organization consolidation can change singleton counts. A source-aware human annotation study should retain candidate singletons, assess false positives and false negatives in complete answers, and report agreement before these curves are interpreted as brand-repertoire estimates. The completed AI audit uses full answers independently of the candidate pool and quantifies agreement under two matching conventions. It does not establish semantic correctness: roles share one model/provider, agreed labels can share errors, and organization mappings are not externally verified. No human reference study was conducted. The separate guarded roster instrument remains unvalidated by that audit.

Second, engine comparisons combine provider, model, response length, output cap, retrieval configuration and backend state. The study supplies one search-enabled September configuration. It therefore cannot identify retrieval as the mechanism causing a smaller observed set. A causal test would vary retrieval within the same model while controlling prompt, output budget and collection window. Results should be described per configuration until such a test is available.

Third, the panel contains 50 deterministically selected English questions collected over several hours. It provides no direct basis for population-wide claims across industries, languages, conversational contexts or future model versions. Exact rarefaction describes this panel, while extrapolating beyond it requires assumptions about stable incidence probabilities and a defined item universe. Repeated runs may share retrieval caches or other backend state. These possibilities were not instrumented. We therefore provide descriptive summaries and make no claim of population-level uncertainty coverage or prospective stopping-rule validation.

\section{Conclusion}
Repeated-query recovery depends on the extraction universe, question panel and collected reference set. On identical September 4 answers, full candidate re-adjudication changes the median ten/fifteen-call recovery range from 92.6\%--95.2\% to 89.5\%--94.7\%. A separate matched five-call comparison produces different system- and subset-specific recovery percentages under a fixed closed-roster matcher. Source-host recovery also depends on whether all returned metadata or visibly referenced entries are counted. These conditional results support reporting finite-sample accumulation and extraction sensitivity together. The blinded AI audit supplies model-reference agreement evidence while leaving human-reference accuracy untested. A new collection with more than ten calls per cell is needed to test the portability of ten-call findings. The evidence does not establish exhaustive discovery, a causal retrieval effect or a universal stopping rule.

\section*{Declarations}
\textbf{Competing interests.} The author is CEO of Rankfor.AI, which supplies AI-visibility measurement and repeated-query brand audits.

\textbf{Funding.} This research received no external funding. Rankfor.AI provided in-kind support for API collection costs.

\textbf{Data and code.} The historical breadth deposit is \href{https://doi.org/10.5281/zenodo.20788142}{Zenodo 10.5281/zenodo.20788142}, published 21 June 2026. Its raw-response and brand-mention files were verified byte-for-byte against the local inputs. The two-wave Category Ownership release is \href{https://doi.org/10.5281/zenodo.22693819}{Zenodo 10.5281/zenodo.22693819}, published 10 September 2026. Its public manifest hashes match both raw collections used in the paired analysis. Historical code and study materials are hosted at \url{https://github.com/Rankfor/rankfor-open/tree/main/research/recommendation-saturation}. The accompanying validation package contains this paper's scripts, source hashes, per-cell results, classifier decisions and AI-review records. The present derived analyses have not yet been deposited as a revised public release.

\textbf{Preprint and overlap.} A prior version is available as \href{https://arxiv.org/abs/2609.05059}{arXiv:2609.05059}, under a different title. Historical corpora are secondary analyses of the author's earlier research. The present revision distinguishes finite-set discovery from the frequency and variance questions of those studies. It reuses the February and September Category Ownership collections; the present estimand is finite-set discovery, rather than that project's brand-frequency and ownership analyses.

\textbf{AI assistance.} Gemini classifiers performed the historical extraction and complete-candidate re-adjudication. Gemini 3.1 Pro Preview performed the two-role full-answer audit and disagreement adjudication described in Methods. OpenAI Codex assisted with manuscript editing, analysis code, numerical checks and source verification. All reviewer personas in the audit were AI-generated roles. No independent human extraction annotation study was conducted.

\appendix
\section{Audit of the historical breadth corpus}
The earlier brief communication described 250 brand-free questions and June 2026 data. The archived raw records instead have timestamps from 23 to 24 February 2026. June 21 is the deposit publication date. Some questions explicitly name competing brands, including ``Salesforce vs HubSpot for B2B marketing.'' The entire historical battery therefore cannot be described as brand-free. These observations were confirmed in the exact public deposit used by the original analysis.

The deposit has 3,750 response slots, including 136 empty strings whose error flag is false. The brand-mention table omits runs with no matched roster brand: 172 slots are absent from that table, comprising 136 empty strings and 36 non-empty answers with no roster match. Grouping the mention table alone consequently drops zero-detection sampling units and produces the previously reported 684 five-run cells.

When all five scheduled slots are retained per cell, treating an empty string as zero observed items for a call-budget sensitivity analysis, there are 750 cells, 13 with empty brand unions. Median observed richness is 4, and 361 of 750 cells contain singletons. Median $A(1)/A(5)$ remains 0.80 over the 737 cells with a non-empty union. This estimand concerns recorded output per scheduled slot; it does not establish whether a blank response reflects a transport, storage or generation failure.

An alternative sensitivity excludes empty answer strings while retaining non-empty answers with no roster match. This yields 737 cells: 707 with five non-empty responses, eight with four, ten with three, five with two and seven with one. Among the 707 complete non-empty cells, median observed richness is 5, singletons occur in 342 cells (48.4\%), and median $A(1)/A(5)$ is 0.80. The historical headline ratio survives these denominator corrections, but the sampling-unit description and counts require revision. This appendix retains the historical mention instrument solely to audit the earlier calculation. Its values are distinct from the guarded-matcher results in the matched-wave analysis and must not be substituted for them.

\section{Four historical deep cells}
The earlier UK supermarket and Nordic care probes contain 24 runs for each of two search-enabled configurations. The collection scripts request GPT-5.4 with web search and Gemini 3.5 Flash with Google Search. The stored recall records use the generic labels \texttt{openai} and \texttt{gemini} and lack per-run timestamps or immutable returned model identifiers. They support a bounded historical comparison, with less complete provenance than the September dataset.

The historical extended dictionaries yield brand-set sizes of 11, 9, 9 and 7 for UK/Gemini, UK/GPT, Nordic/Gemini and Nordic/GPT, respectively. All four have $Q_1=0$. Expected ten-run recovery of the observed 24-run brand sets is 95.4\%, 100.0\% when rounded, 95.5\% and 92.0\%, respectively. The earlier 91\% lower endpoint is not reproduced by the stored revised extraction. Zero singletons describe recurrence among the observed extracted items, without certifying absence of unseen items.

The corresponding source-host sets contain 48, 18, 97 and 29 hosts, with $Q_1=18,6,41,12$ and Chao2 values 67.4, 21.5, 164.1 and 40.5. A single sampled run contains 23.1\%, 16.0\%, 18.3\% and 19.5\% of the observed 24-run host sets. This supports different brand and source accumulation within these four cells. It supplies no universal rule about search-enabled systems.

\end{document}